\documentstyle[12pt,epsf]{article}
 \newcommand{\insertplot}[5]{\begin{figure}
 \hfill\hbox to 0.05in{\vbox to #5in{\vfill
 \inputplot{#1}{#4}{#5}}\hfill}
 \hfill\vspace{-.1in}
 \caption{#2}\label{#3}
 \end{figure}}
 \newcommand{\inputplot}[3]{
 \special{ps: plotfile #1}
}

\begin{document}

\title{SU(3) Einstein-Skyrme Solitons and Black Holes}
\vspace{1.5truecm}
\author{
{\bf Burkhard Kleihaus$^1$, Jutta Kunz$^{1,2}$ and Abha Sood$^1$}\\
$^1$Fachbereich Physik, Universit\"at Oldenburg, Postfach 2503\\
D-26111 Oldenburg, Germany\\
$^2$
Instituut voor Theoretische Fysica, Rijksuniversiteit te Utrecht\\
NL-3508 TA Utrecht, The Netherlands}

\date{March 13, 1995}

\maketitle
\vspace{1.0truecm}

\begin{abstract}
In the SU(3) Einstein-Skyrme system
static spherically symmetric particle-like solutions
and black holes exist for both the SU(2) and the SO(3) embedding.
The SO(3) embedding leads to new particle-like solutions
and black holes, sharing many features with the SU(2) solutions.
In particular, there are always two branches of solutions,
forming a cusp at a critical coupling constant.
The regular SO(3) solutions have even topological charge $B$.
The mass of the $B=2$ SO(3) solutions
is less than twice the mass of the $B=1$ SU(2) solutions.
We conjecture, that the lowest SO(3) branches
correspond to stable particle-like solutions and
stable black holes.
\end{abstract}
\vfill
\noindent {Utrecht-Preprint THU-95/6; hep-th/9503087} \hfill\break
\vfill\eject

\section{Introduction}

The SU(2) Einstein-Skyrme system possesses
two different types of black hole solutions with the same mass
for a certain range of coupling constants \cite{luck,droz,bizon}.
These are the Schwarzschild black holes
with vanishing chiral fields
and the SU(2) Einstein-Skyrme black holes with ``chiral hair''.
For a given event horizon, there are always two branches of
black hole solutions with ``chiral hair''
which form a cusp at a critical coupling constant
(dependent on the event horizon)
\cite{bizon}.
The black hole solutions on the lower branches are
linearly stable \cite{heusler,bizon}.
Thus for a certain range of masses the system possesses
two distinct classically stable types of black holes,
providing a counterexample to the ``no-hair conjecture''
for black holes.

Beside the black hole solutions
the SU(2) Einstein-Skyrme system also possesses
two branches of regular solutions,
which merge at a critical coupling constant \cite{bizon}.
Again, the solutions on the lower branch are
classically stable \cite{bizon,heusler},
they correspond to solitons.
For zero coupling
the lower branch starts from the SU(2) skyrmion solution
in flat space \cite{skyrme,anw},
while the upper branch ends in the
lowest SU(2) Einstein-Yang-Mills sphaleron solution \cite{bm}
(after scaling the radial coordinate \cite{bizon}).

Here we consider particle-like solutions and black holes
in the SU(3) Einstein-Skyrme system.
In the SU(3) Skyrme model in flat space
regular static spherically symmetric solutions
exist for both subgroups SU(2) and SO(3) \cite{bala}.
The SU(2) embedding yields as lowest solution
the stable SU(2) skyrmion with topological number $B=1$.
The SO(3) embedding yields as lowest solution
a stable solution with topological number $B=2$,
whose mass is lower than twice the mass of the SU(2) skyrmion
\cite{bala}.
These SO(3) solutions are recovered in
the SU(3) Einstein-Skyrme system for vanishing coupling.
They form the starting point of our investigation.
We then vary the coupling constant
to construct the whole family of particle-like SO(3) solutions,
consisting of a lower and an upper branch.
As endpoint on the upper branch we obtain an SO(3) sphaleron
solution (after scaling the radial coordinate).

Analogous to the SU(2) Einstein-Skyrme system,
asymptotically flat SO(3) black hole solutions
emerge from the regular solutions by requiring regularity
at a finite event horizon.
For a given horizon,
two branches of SO(3) black holes with ``chiral hair''
exist below a critical coupling constant
(dependent on the horizon).
We argue, that the SO(3) black hole solutions
on the lower branch are classically stable.

\section{SU(3) Einstein-Skyrme Equations of Motion}

We consider the SU(3) Einstein-Skyrme action
\begin{equation}
S=S_G+S_M=\int L_G \sqrt{-g} d^4x + \int L_M \sqrt{-g} d^4x
\   \end{equation}
with
\begin{equation}
L_G=\frac{1}{16\pi G}R
\ , \end{equation}
and the matter Lagrangian
for the chiral SU(3) field $U$
\begin{equation}
L_M = \frac{1}{4} f^2 {\rm Tr} (A_\mu A^\mu)
    +\frac{1}{32e^2} {\rm Tr} (F_{\mu\nu} F^{\mu\nu})
\ , \end{equation}
where
\begin{equation}
A_\mu = U^\dagger \nabla_\mu U = U^\dagger \partial_\mu U
\   \end{equation}
and
\begin{equation}
F_{\mu\nu}= [A_\mu,A_\nu]
\ , \end{equation}
and $f$ and $e$ are coupling constants.
Variation of the action eq.~(1) with respect to the metric
$g_{\mu\nu}$ and the chiral field $U$
leads to the Einstein equations and the matter field equations.

To construct static spherically symmetric solitons and black holes
we employ Schwarz\-schild-like coordinates and adopt
the spherically symmetric metric
\begin{equation}
ds^2=g_{\mu\nu}dx^\mu dx^\nu=
  -A^2N dt^2 + N^{-1} dr^2 + r^2 (d\theta^2 + \sin^2\theta d\phi^2)
\ , \end{equation}
with
\begin{equation}
N=1-\frac{2m}{r}
\ . \end{equation}
Generalized spherical symmetry for the chiral field is realized
by embedding the SU(2) or the SO(3) generators $T_i$ in SU(3).
In the SU(2)-embedding
$\vec T = \frac{1}{2} (\lambda_1, \lambda_2, \lambda_3)$,
and the ansatz for the chiral field is
\begin{equation}
U =
    \left( \begin{array}{cc}
 \cos{\xi}+ i \tau_r \sin{\xi} & 0 \\
 0 & 1 \end{array} \right)
\ , \label{su2} \end{equation}
with the SU(2) Pauli matrices $\tau_i$ and
$\tau_r = (\tau_1,\tau_2,\tau_3)\cdot \vec e_r$.
(The general ansatz for $U$ includes a factor of
$\exp({i\lambda_8 \zeta})$,
however $\zeta=0$ for the lowest solution
 \cite{bala}.)
In the SO(3)-embedding
$\vec T = (\lambda_7, -\lambda_5, \lambda_2)$,
and the ansatz for the chiral field is
\begin{equation}
U=\exp\Bigl(i\Lambda_r \chi + i( \Lambda_r^2-\frac{2}{3})\phi\Bigr)
\  \label{so3} \end{equation}
with
$\Lambda_r = (\lambda_7,-\lambda_5,\lambda_2)\cdot \vec e_r$.
An equivalent expression for the chiral field is
\begin{equation}
U=\exp(i\psi) + i \Lambda_r \exp(-i\frac{\psi}{2})\sin\chi
 + \Lambda_r^2 \Bigr(
                 \exp(-i\frac{\psi}{2}) \cos\chi - \exp(i\psi) \Bigr)
\ , \end{equation}
with $\phi=-3\psi/2$.

The SU(2)-embedding, eq.~(\ref{su2}),
leads to the well studied SU(2) Einstein-Skyrme equations
\cite{droz,bizon}.
To obtain the SU(3) Einstein-Skyrme equations
for the SO(3)-embed\-ding, eq.~(\ref{so3}),
we also employ the $tt$ and $rr$
components of the Einstein equations,
yielding
\begin{eqnarray}
\mu'&=&\alpha \Biggl(
  \frac{1}{2} x^2 N \Bigl( \chi'^2 +\frac {1}{3} \phi'^2 \Bigr)
  +2 ( 1-\cos\phi \cos \chi )
\nonumber\\
 & &\phantom{ \alpha \Biggl( }
 +\frac{1}{2} N (1-\cos\phi \cos \chi ) (\chi'^2 + \phi'^2)
  + N \sin\phi \sin \chi \phi' \chi'
\nonumber\\
 & &\phantom{ \alpha \Biggl( }
  +\frac{3}{2 x^2} \sin^2 \phi \sin^2 \chi
  +\frac{1}{2 x^2} ( 1-\cos\phi\cos\chi)^2  \Biggr)
\ , \end{eqnarray}
and
\begin{eqnarray}
 A'&=&\alpha \Biggl(
   x \Bigl(\chi'^2 +\frac {1}{3} \phi'^2 \Bigr)
\nonumber\\
 & &\phantom{ \alpha \Biggl( }
 +\frac{1}{x} \Bigl( (1-\cos\phi \cos \chi ) (\chi'^2 + \phi'^2)
  + 2 \sin\phi \sin \chi \phi' \chi' \Bigr)  \Biggr) A
\ , \label{eqa} \end{eqnarray}
where we have introduced the dimensionless mass function
\begin{equation}
\mu=efm
\   \end{equation}
and the dimensionless coordinate
$x=efr$,
the prime indicates the derivative
with respect to $x$,
and
\begin{equation}
\alpha = 4 \pi G f^2  = 4 \pi \bigl( \frac{f}{m_{\rm Pl}} \bigr)^2
\   \end{equation}
is a dimensionless coupling constant.
For the matter functions we obtain the equations
\begin{eqnarray}
\lefteqn {\Biggl( AN \Bigl(  (1+x^2 - \cos\phi \cos\chi) \chi'
       +\sin \phi \sin \chi \phi' \Bigr) \Biggr)'
  \phantom{extend a bit more} }
\nonumber\\
 &=& A \Biggl( \cos\phi \sin\chi \Bigl(
      2 + \frac{N}{2}(\chi'^2 +\phi'^2)
    + \frac{1}{x^2} (1-\cos\phi \cos\chi) \Bigr)
\nonumber\\
 & & \phantom{A \Biggl(}
  + \sin\phi \cos\chi \Bigl( N \phi' \chi'
     +\frac{3}{x^2} \sin\phi \sin\chi \Bigr) \Biggr)
\ , \end{eqnarray}
and
\begin{eqnarray}
\lefteqn {\Biggl( AN \Bigl(  (1+\frac{1}{3} x^2
                             - \cos\phi \cos\chi) \phi'
       +\sin \phi \sin \chi \chi' \Bigr) \Biggr)'
  \phantom{extend a bit more} }
\nonumber\\
 &=& A \Biggl( \sin\phi \cos\chi \Bigl(
      2 + \frac{N}{2}(\chi'^2 +\phi'^2)
    + \frac{1}{x^2} (1-\cos\phi \cos\chi) \Bigr)
\nonumber\\
 & & \phantom{A \Biggl(}
  + \cos\phi \sin\chi \Bigl( N \phi' \chi'
     +\frac{3}{x^2} \sin\phi \sin\chi \Bigr) \Biggr)
\ . \end{eqnarray}
With help of eq.~(\ref{eqa}) the metric function $A$
can be eliminated from the matter equations.

The Einstein-Skyrme system has a topological current
\begin{equation}
B^\mu = -\frac{1}{\sqrt{-g}} \frac{1}{24\pi^2}
\epsilon^{\mu \nu \alpha \beta} {\rm Tr} A_\nu A_\alpha A_\beta
\   \end{equation}
which is covariantly conserved,
yielding the topological charge
\begin{equation}
B = \int \sqrt{-g} B^0 d^3 x
 = -\frac{1}{2 \pi^2} \int
 \Bigl( (1- \cos\phi \cos\chi) \chi' + \sin \phi \sin \chi \phi' \Bigr)
  dx \sin \theta  d\theta d\phi
\ . \end{equation}

\section{Regular Solutions}

Let us consider the regular particle-like solutions
of the SU(3) Einstein-Skyrme system.
Requiring asymptotically flat solutions implies
that the metric functions $A$ and $\mu$ both
approach a constant at infinity,
and that the chiral functions
approach an integer multiple of $\pi$.
We here adopt
\begin{equation}
A(\infty)=1
\ , \end{equation}
thus fixing the time coordinate, and
\begin{equation}
\chi(\infty)=0 \ , \quad \phi(\infty)=0
\ . \end{equation}
At the origin regularity of the solutions requires
\begin{equation}
\mu(0)=0
\ , \end{equation}
and that the chiral functions
again approach an integer multiple of $\pi$.
The lowest non-trivial solution has \cite{bala}
\begin{equation}
\chi(0)=\pi \ , \quad \phi(0)=\pi
\ , \label{bc1} \end{equation}
and the choice $\chi(0)=\pi$, $\phi(0)=-\pi$
leads to a degenerate solution.
The boundary conditions
\begin{equation}
\chi(0)=2\pi \ , \quad \phi(0)=0
\ , \label{bc2} \end{equation}
lead to the known SU(2) solutions
corresponding to the modified coupling constant
$\tilde \alpha = 4 \alpha$
with $\chi= 2\xi$ and $\phi=0$.
(Thus for the boundary conditions (\ref{bc2})
the SO(3) masses for the coupling constant $\alpha$
are four times as big as the SU(2) masses
for the coupling constant $\tilde \alpha$,
and the critical SO(3) coupling constant is one forth
of the critical SU(2) coupling constant.)

The topological charge of the solutions \cite{bala}
\begin{equation}
B= - \left. \frac{2}{\pi} (\chi - \sin \chi \cos \phi )
 \right|^{\infty}_0
\   \end{equation}
depends only on the boundary conditions.
For the boundary conditions (\ref{bc1}) and (\ref{bc2})
the topological charge is
$B=2$ and $B=4$, respectively.
In the following we present the numerical
solutions for the SO(3) embedding
with boundary conditions (\ref{bc1}),
corresponding to $B=2$ solutions,
since the solutions of the SU(2) embedding reproduce
the known SU(2) results.

The regular SO(3) Einstein-Skyrme solutions
and the regular SU(2) Einstein-Skyrme solutions
have many features in common.
The equations depend only on the coupling constant $\alpha=4\pi G f^2$,
whose variation can be considered in two different ways,
either as changing $G$ and keeping $f$ constant, or vice versa.
This results in the occurrence of two distinct branches
of solutions, which exist for a finite range
of coupling constants $[0,\alpha_c]$.
The branches merge in a cusp at $\alpha_c$
and beyond this critical coupling no solutions exist.
The value of $\alpha_c$ depends on the embedding
and on the boundary conditions.

\begin{figure}
\centering
\epsfysize=11cm
\mbox{\epsffile{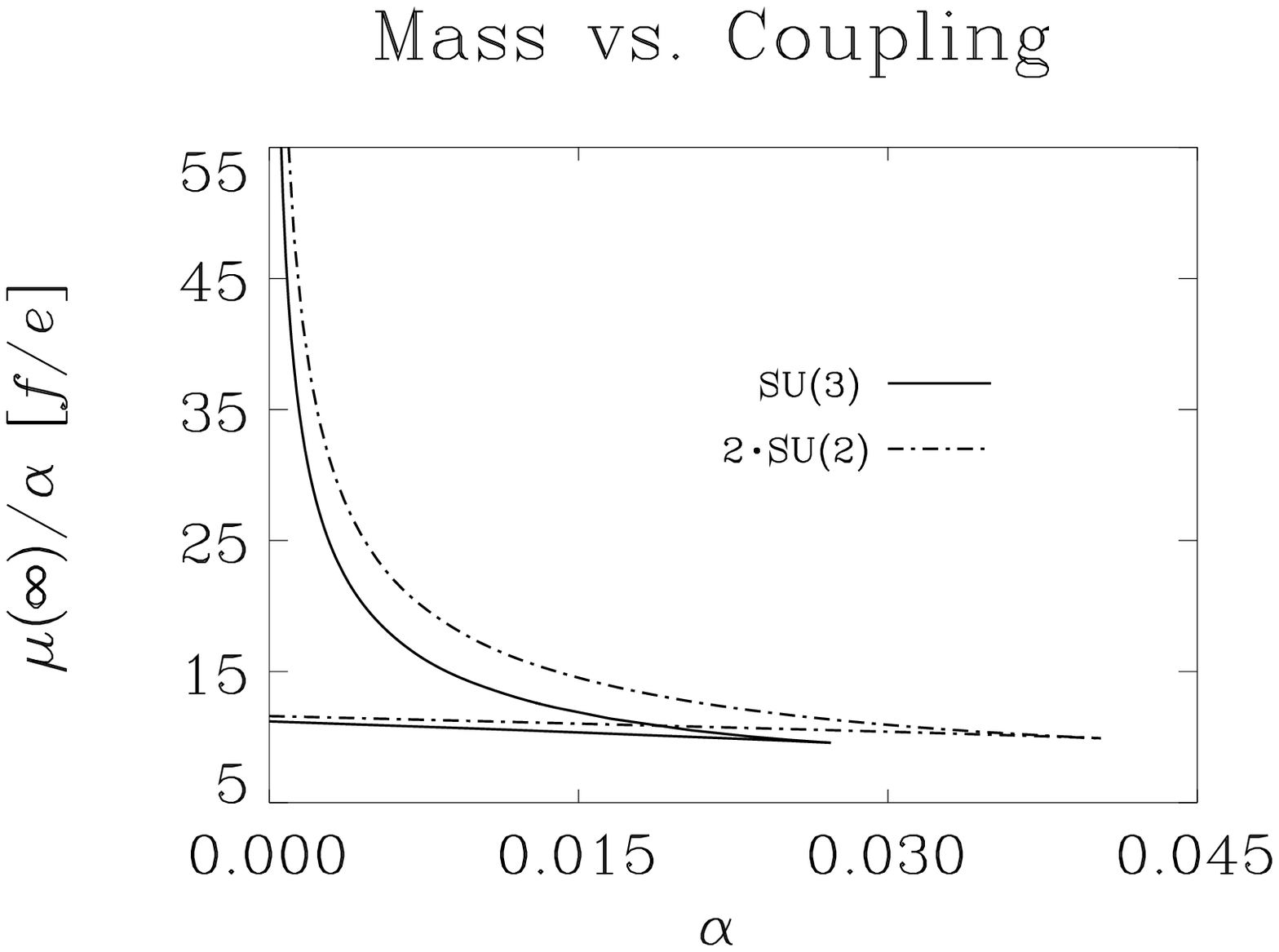}}
\caption{\label{regen}
The masses of the regular $B=2$ SO(3) solutions
and twice the masses of the regular $B=1$ SU(2) solutions
are shown as a function of the coupling constant $\alpha$.
Multiplication with $4\pi$ gives the ADM mass.}
\end{figure}

In Fig.~\ref{regen} we demonstrate this feature
by presenting the mass of the $B=1$ SU(2) solutions
and the $B=2$ SO(3) solutions along both branches.
The mass of the SO(3) solutions is always lower than
twice the mass of the SU(2) solutions.
The ADM mass
\begin{equation}
m_{\rm ADM}= \mu(\infty) \frac{4 \pi }{\alpha} \frac{f}{e}
\end{equation}
can be read off the figure,
and the critical SO(3) coupling is $\alpha_c = 0.02728$,
while the critical SU(2) coupling is $\alpha_c=0.04038$
\cite{droz,bizon}.

Considering the limit $\alpha \rightarrow 0$,
the solutions on the lower branch reduce to those of the Skyrme model
in flat space with $\mu=0$.
We therefore refer to this branch as the skyrmion branch.
In the limit $\alpha \rightarrow 0$
the solutions on the upper branch shrink to zero size,
while their mass diverges.
This is, however, only due to the particular
choice of variables, as observed by Bizon and Chmaj \cite{bizon}
for SU(2).
With the coordinate transformation
$\tilde x = x/\sqrt{\alpha}$ \cite{bizon}
the equations for the metric and matter functions
become identical to those of the static spherically symmetric magnetic
Einstein-Yang-Mills system
in the limit $\alpha \rightarrow 0$,
where sphaleron solutions exist \cite{kuenzle,we}.
We therefore refer to the upper branch as sphaleron branch.
Here we note only, that the mass
of the lowest SO(3) sphaleron
is $\tilde \mu(\infty)=1.3078$ \cite{we},
while the mass of the lowest SU(2) sphaleron \cite{bm}
is $\tilde \mu(\infty)=0.8286$,
and $m_{\rm ADM}= \tilde \mu(\infty) \sqrt{4 \pi} m_{\rm Pl}/e$.

\begin{figure}
\centering
\epsfysize=11cm
\mbox{\epsffile{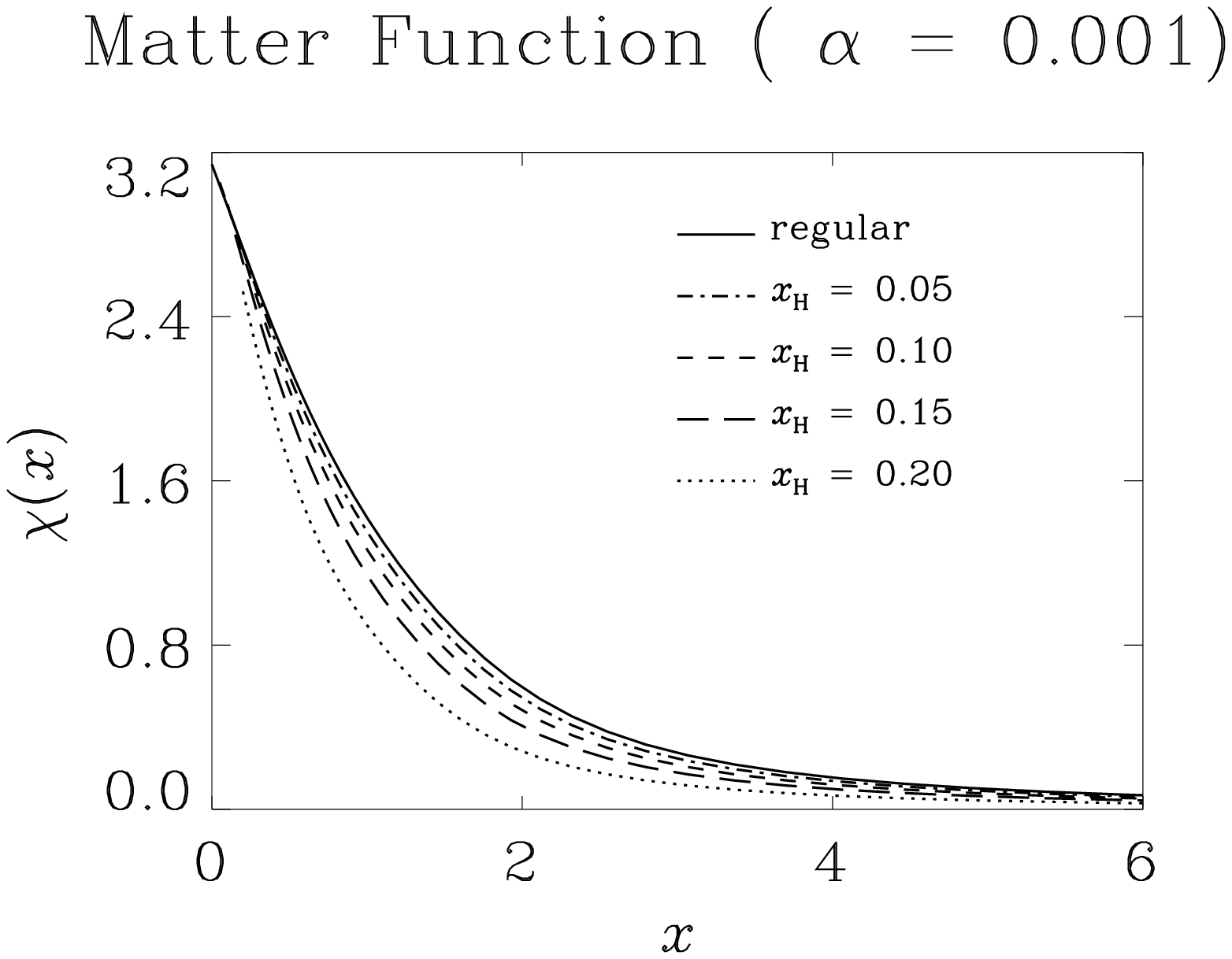}}
\caption{\label{chi}
The function $\chi(x)$
is shown for the regular solution and for the
black hole solutions with horizons
$x_{\rm H}=0.05$, 0.10, 0.15 and 0.20
as a function of $x$
for the coupling $\alpha=0.001$.}
\end{figure}

\begin{figure}
\centering
\epsfysize=11cm
\mbox{\epsffile{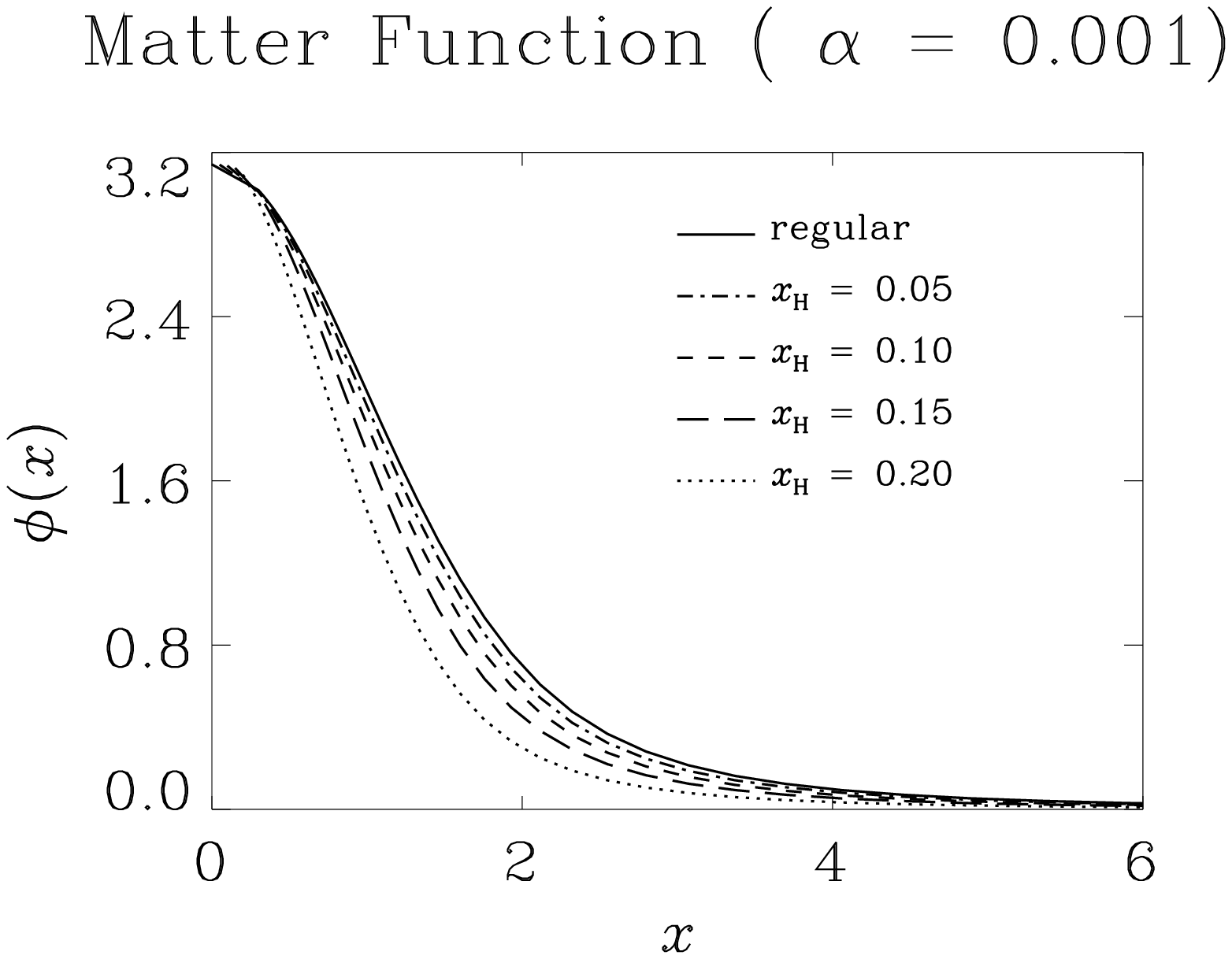}}
\caption{\label{phi}
The function $\phi(x)$
is shown for the regular solution and for the
black hole solutions with horizons
$x_{\rm H}=0.05$, 0.10, 0.15 and 0.20
as a function of $x$
for the coupling $\alpha=0.001$.}
\end{figure}

\begin{figure}
\centering
\epsfysize=11cm
\mbox{\epsffile{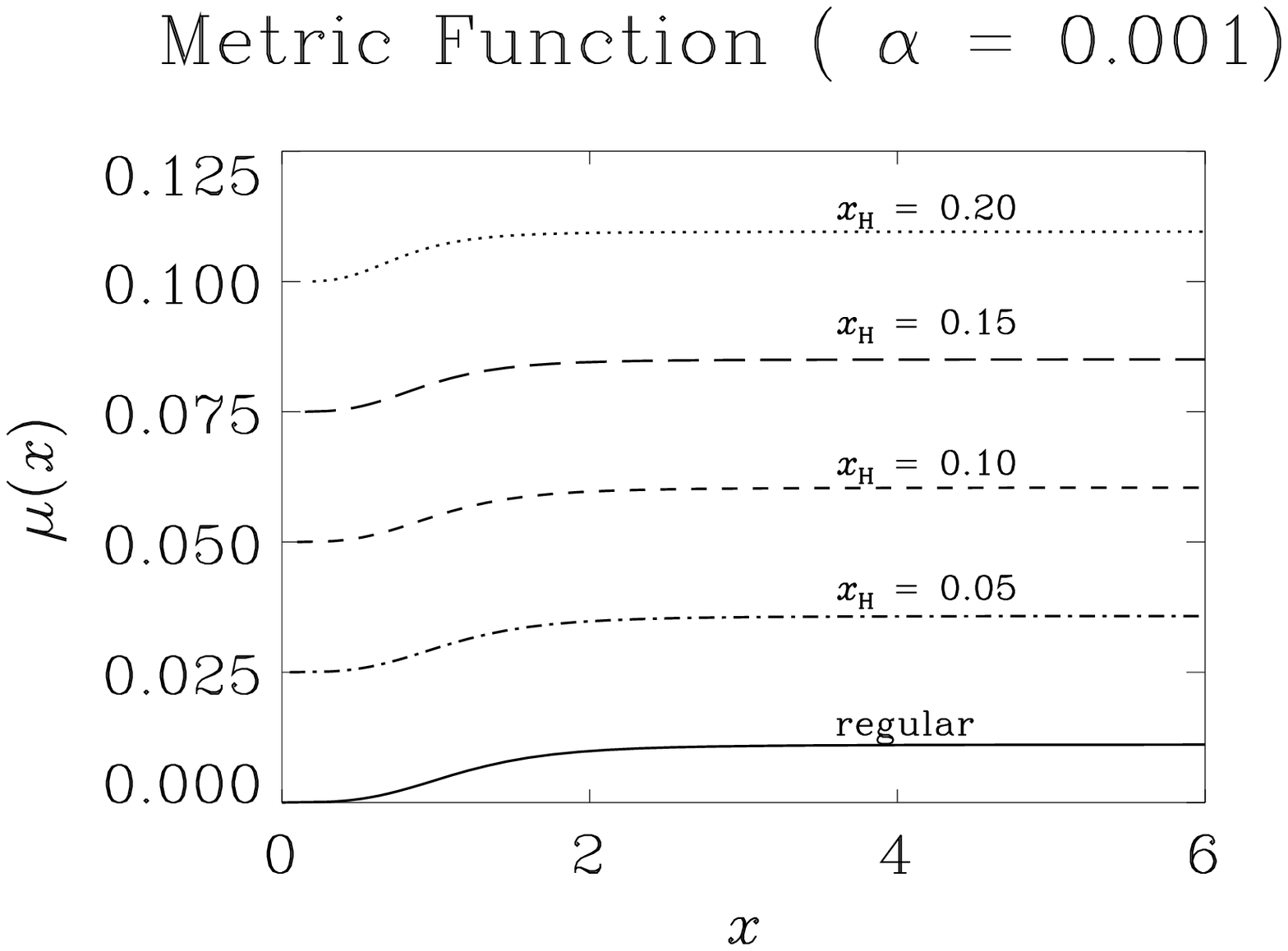}}
\caption{\label{mu}
The function $\mu(x)$
is shown for the regular solution and for the
black hole solutions with horizons
$x_{\rm H}=0.05$, 0.10, 0.15 and 0.20
as a function of $x$
for the coupling $\alpha=0.001$.}
\end{figure}

\begin{figure}
\centering
\epsfysize=11cm
\mbox{\epsffile{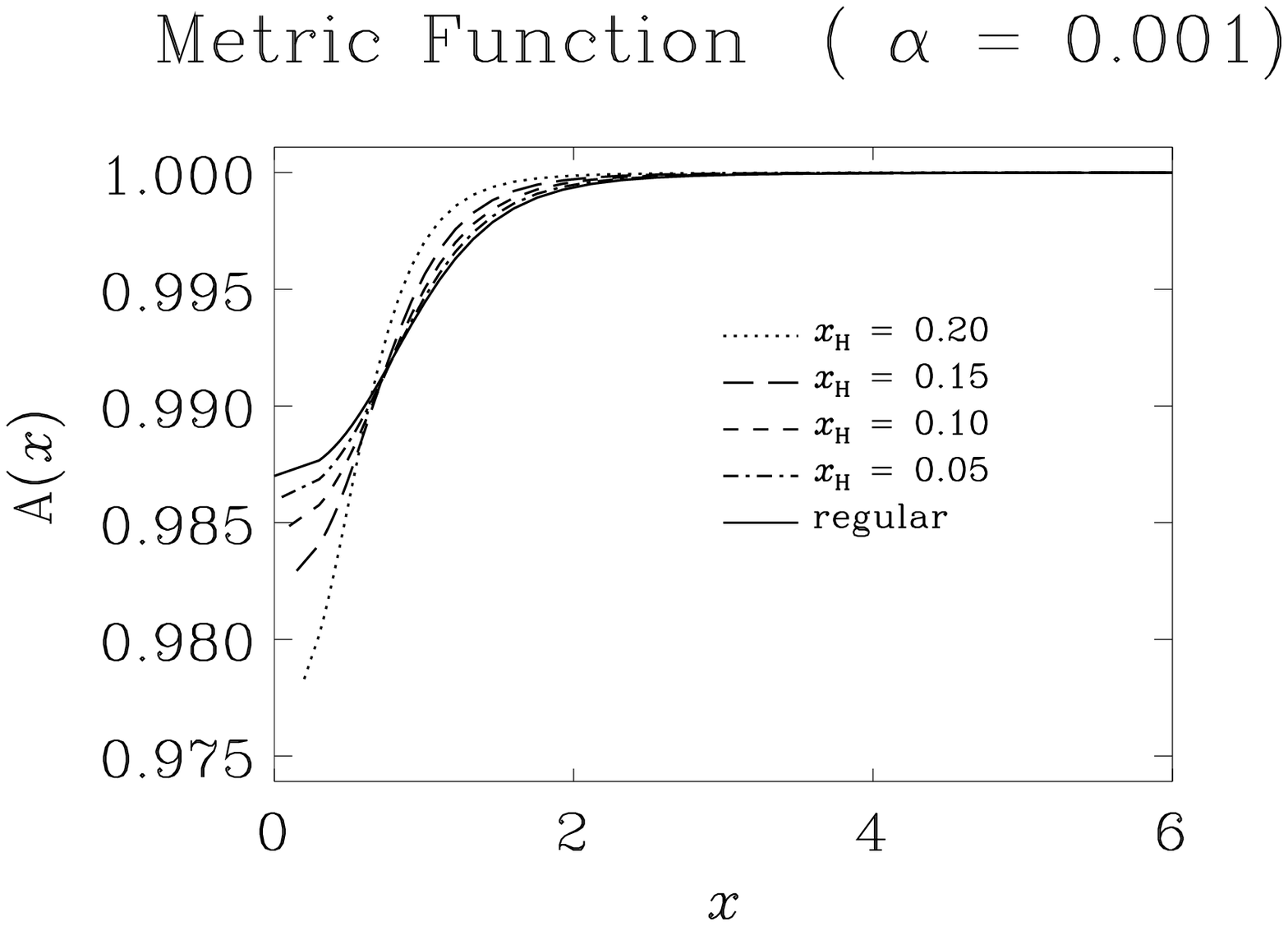}}
\caption{\label{a}
The function $A(x)$
is shown for the regular solution and for the
black hole solutions with horizons
$x_{\rm H}=0.05$, 0.10, 0.15 and 0.20
as a function of $x$
for the coupling $\alpha=0.001$.}
\end{figure}

Let us now consider the family of regular SO(3) solutions
obtained by first increasing $\alpha$
along the skyrmion branch from zero
up to the critical coupling $\alpha_c$,
and then decreasing $\alpha$ again along
the sphaleron branch back to zero.
Starting from the SO(3) skyrmion solution in flat space,
the regular solutions change continuously,
and reach finally the SO(3) sphaleron solution \cite{we},
where they shrink to zero size
(with respect to the coordinate $x$).
Along this path, the solutions can be characterized,
for instance, by the derivative of $\chi$ at the origin,
which changes monotonically.
As an example we show the radial functions for the regular
SO(3) solution along the skyrmion branch
for $\alpha = 0.001$ in Figs.~(\ref{chi})-(\ref{a}).
Further details will be given elsewhere \cite{we}.

\section{Black Hole Solutions}

We now turn to the black hole solutions of the
SU(3) Einstein-Skyrme system.
Imposing again the condition of asymptotic flatness,
the black hole solutions satisfy the same
boundary conditions at infinity
as the regular solutions.
The existence of a regular event horizon at $x_{\rm H}$
requires
\begin{equation}
\mu(x_{\rm H})= \frac{x_{\rm H}}{2}
\ , \end{equation}
and $A(x_{\rm H}) < \infty $,
and the chiral functions must satisfy at the horizon $x_{\rm H}$
\begin{eqnarray}
 N' \Bigl(  (1+x^2 - \cos\phi \cos\chi) \chi'
       +\sin \phi \sin \chi \phi' \Bigr)
  \phantom{extend a bit more}
\nonumber\\
  =   \cos\phi \sin\chi \Bigl( 2
    + \frac{1}{x^2} (1-\cos\phi \cos\chi) \Bigr)
     +\frac{3}{2x^2} \sin^2\phi \sin2\chi
\ , \\
 N' \Bigl(  (1+\frac{1}{3}x^2 - \cos\phi \cos\chi) \phi'
       +\sin \phi \sin \chi \chi' \Bigr)
  \phantom{extend a bit more}
\nonumber\\
  =   \sin\phi \cos\chi \Bigl( 2
    + \frac{1}{x^2} (1-\cos\phi \cos\chi) \Bigr)
     +\frac{3}{2x^2} \sin2\phi \sin^2\chi
\ . \end{eqnarray}

\begin{figure}
\centering
\epsfysize=11cm
\mbox{\epsffile{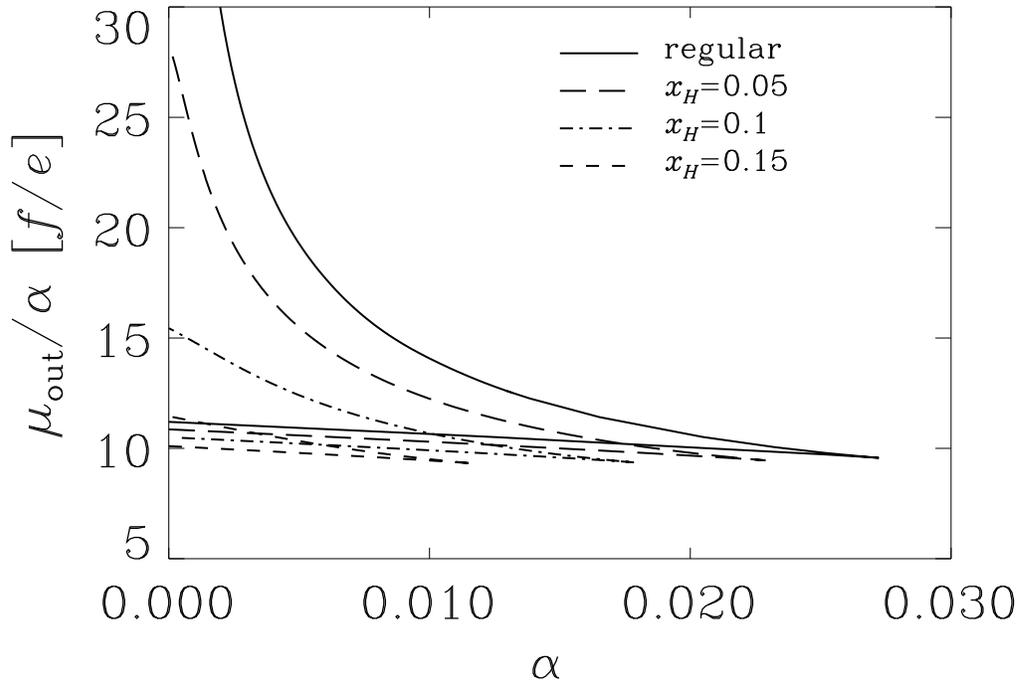}}
\caption{\label{bhen}
The masses of the regular SO(3) solutions
and the mass fractions outside the horizon $\mu_{\rm out}$
of the SO(3) black hole solutions
are shown as a function of the coupling constant $\alpha$.
When the mass fractions within the horizon, $x_{\rm H}/2$, are added,
multiplication with $4\pi$ gives the ADM masses.}
\end{figure}

Again, the SO(3) Einstein-Skyrme black holes
and the SU(2) Einstein-Skyrme black holes
have many features in common.
For a given horizon $x_{\rm H}$
there are two distinct branches of solutions,
which exist for a finite range
of coupling constants $[0,\alpha_c(x_{\rm H})]$.
The critical coupling now depends on the horizon $x_{\rm H}$.
In Fig.~\ref{bhen} we exhibit the masses of
the SO(3) black holes in terms of the mass fractions outside
the horizon, $\mu_{\rm out}$, defined via
\begin{equation}
m_{\rm ADM} = \Bigl( \frac{x_{\rm H}}{2} + \mu_{\rm out} \Bigr)
  \frac{4 \pi }{\alpha} \frac{f}{e}
= \mu(\infty) \frac{4 \pi }{\alpha} \frac{f}{e}
\ . \end{equation}
For $x_{\rm H} \rightarrow 0$ the black hole solutions
approach the regular solutions.
With increasing horizon $x_{\rm H}$
the critical coupling $\alpha_c(x_{\rm H})$
decreases from the critical value of the regular solutions $\alpha_c$.

The families of SO(3) black hole solutions
for various values of the horizon $x_{\rm H}$
change continuously,
when $\alpha$ is increased from zero
along the skyrmion branch
up to the critical coupling $\alpha_c(x_{\rm H})$,
and then decreased back to zero
along the sphaleron branch.
At zero coupling on the skyrmion branch
the solutions correspond to
Schwarz\-schild black holes with ``chiral hair''.
As examples we show the radial functions for the
SO(3) black hole solutions with the horizons
$x_{\rm H}=0.05$, 0.10, 0.15 and 0.20
along the skyrmion branch
for $\alpha = 0.001$ in Figs.~(\ref{chi})-(\ref{a}).
Further details of these solutions
as well as the SO(3) Einstein-Yang-Mills black holes
will be given elsewhere \cite{we}.

\section{Conclusion}

Like the regular SU(2) solutions with topological number $B=1$,
the regular SO(3) solutions with topological number $B=2$
form two branches,
a skyrmion branch and a sphaleron branch,
existing below a critical coupling.
The corresponding critical SU(2) and SO(3) couplings are
$\alpha_c=0.04038$ \cite{droz,bizon} and $\alpha_c = 0.02728$,
respectively.
(Thus the critical coupling of the B=2 solutions
in the SO(3) embedding
is considerably larger than that of the B=2 solutions
in the SU(2) embedding \cite{bizon}.)
For vanishing coupling,
the lower branch of the regular SO(3) solutions
starts at the $B=2$ skyrmion in flat space \cite{bala},
and the upper branch ends in the lowest SO(3)
Einstein-Yang-Mills sphaleron \cite{kuenzle,we}
(after scaling the radial coordinate).

For a given coupling constant,
on both branches
the mass of the regular SO(3) solution with $B=2$
is smaller than twice the mass of the SU(2)
solution with $B=1$.
The $B=1$ SU(2) and the $B=2$ SO(3) skyrmion solutions
in flat space are stable \cite{bala},
and the full SU(2) skyrmion branch is stable \cite{heusler,bizon}.
We conclude, that the full SO(3) skyrmion branch
is also stable.

General arguments from catastrophe theory imply
a change of stability at the critical coupling constant.
(This fact is also known for skyrmions coupled to vector bosons
in flat space \cite{eilam,bk,bsk}.)
For the SU(2) case it was shown explicitly,
that one eigenvalue passes zero at the critical coupling \cite{bizon}.
The SU(2) solutions on the sphaleron branch
then possess one unstable mode \cite{bizon}.
(Note, that the
lowest SU(2) Einstein-Yang-Mills sphaleron,
has two unstable modes \cite{strau,volkov1,lav}.)
Therefore we argue, that a change of stability
takes place at the cusp for the SO(3) solutions as well.
The SO(3) solutions on the sphaleron branch
then should possess one unstable mode.

For a given value of the event horizon,
the SO(3) black hole solutions also form two branches,
a skyrmion branch and a sphaleron branch,
with a bifurcation point at a critical coupling.
This critical coupling depends on the horizon,
and decreases with increasing horizon.
The $B=1$ SU(2) skyrmion branch is known to be stable,
while the sphaleron branch is unstable
with one unstable mode \cite{heusler,bizon}.
Invoking arguments from catstrophe theory again \cite{maeda},
we conjecture, that the $B=2$ SO(3) skyrmion branch
is stable, too, while the sphaleron branch is unstable
with one unstable mode, acquired at the critical coupling.
Since the SO(3) Einstein-Skyrme system also contains
the Schwarz\-schild black holes,
there are then for a certain range of coupling constants
(depending on the event horizon),
three black hole solutions,
two of which are stable:
the Schwarz\-schild solution and the black hole
solution with ``chiral hair'' on the skyrmion branch.
This then provides an additional counterexample
to the ``no-hair conjecture''.

The SU(2) Einstein-Skyrme model has a rich spectrum
of particle-like solutions and black holes \cite{bizon},
on the one hand corresponding to solutions
in higher topological sectors
and on the other hand corresponding to solutions
related to the excited SU(2) Einstein-Yang-Mills sphalerons
\cite{bm} and black holes
\cite{volkov,bizon1}.
For the SO(3) Einstein-Skyrme system
we expect a similarly rich spectrum,
which is presently under investigation.
Solutions in higher topological sectors
certainly exist,
and so do excited SO(3) Einstein-Yang-Mills sphalerons
\cite{kuenzle,we}.

{\sl Acknowledgement} We gratefully acknowledge discussions
with M. Volkov.

\newpage

\newpage

\newpage

\newpage

\newpage

\newpage

\end{document}